\documentclass[11pt]{article} 

\usepackage[ansinew]{inputenc}
\usepackage{amsmath, amssymb, graphics, amsthm}
\usepackage{epsfig}
\usepackage{color}
\usepackage{undertilde}
\usepackage{fancyhdr}

\newcommand{\m}{\mbox{}}

\makeatletter
\@addtoreset{equation}{section}
\makeatother

\newcommand{\be}{\begin{equation}}
\newcommand{\ee}{\end{equation}}
\newcommand{\ba}{\begin{eqnarray}}
\newcommand{\ea}{\end{eqnarray}}

\title{{\sf New Variables for Classical and Quantum Gravity}\\
{\sf in all Dimensions IV. Matter Coupling}} 
\author{
{\sf N. Bodendorfer}$^{1,2}$\thanks{{\sf 
norbert.bodendorfer@gravity.fau.de}},
{\sf T. Thiemann}$^{1,3}$\thanks{{\sf 
thomas.thiemann@gravity.fau.de,
tthiemann@perimeterinstitute.ca}},
{\sf A. Thurn}$^1$\thanks{{\sf 
andreas.thurn@gravity.fau.de}}\\
\\
{\sf $^1$ Inst. for Theoretical Physics III, FAU Erlangen -- N\"urnberg,}\\
{\sf Staudtstr. 7, 91058 Erlangen, Germany}\\
\\
{\sf $^2$ Institute for
Gravitation and the Cosmos \& Physics
  Department,}\\
{\sf   Penn State, University Park, PA 16802, U.S.A.}\\
\\
{\sf $^3$ Perimeter Institute for Theoretical Physics,}\\ 
{\sf 31 Caroline Street N, Waterloo, ON N2L 2Y5, Canada}
}
\date{{\small\sf \today}}

\begin{document} 

\maketitle

{\sf

\begin{abstract}
We employ the techniques introduced in the companion papers \cite{BTTI,BTTII,BTTIII} to derive a connection formulation of Lorentzian General Relativity coupled to Dirac fermions in dimensions 
$D+1\geq 3$ with compact gauge group. The technique that accomplishes that is similar 
to the one that has been 
introduced in $3+1$ dimensions already:  First one performs a canonical analysis of Lorentzian General Relativity using the time gauge and then introduces an extension of the phase space 
analogous to the one employed in  \cite{BTTI} to obtain a connection theory with SO$(D+1)$ as the internal gauge 
group subject to additional constraints. The success of this method rests heavily on the strong similarity of the Lorentzian and Euclidean Clifford algebras. A quantisation of the Hamiltonian constraint is provided.
\end{abstract}

}

\newpage

\section{Introduction}

In the previous papers of this series \cite{BTTI,BTTII,BTTIII} we introduced a new connection 
formulation of vacuum General Relativity with compact gauge group in any spacetime 
dimension $D+1\ge 3$. In this paper we consider coupling of this theory to matter.
Here it will suffice to consider fermionic matter because gauge bosons can be coupled 
in the same way as it has been done in $3+1$ dimensions already \cite{ThiemannQSD5}: Scanning 
through the details of \cite{ThiemannQSD5} one realises that nothing depends substantially 
on $D=3$ and since the background independent LQG representation for theories
of connections as well as its uniqueness generalises to any dimension, we can consider
the gauge boson sector as treated already.     

Our starting point is the standard canonical treatment of fermions coupled to General Relativity:\\ 
To the best of our knowledge, Kibble \cite{KibbleCanonicalVariablesFor} was the first to consider the canonical formulation of fermions coupled to vierbein gravity. The classical coupling of fermions 
to the new  variables \cite{AshtekarNewVariablesFor} was provided in  \cite{AshtekarInclusionOfMatter}. Since then, several papers appeared debating issues arising when including fermions. Among others, the role of the Immirzi parameter \cite{PerezPhysicalEffectsOf}, the appearance of torsion \cite{BojowaldCanonicalGravityWith, MercuriFermionsInThe} and the correct form of the Holst modification \cite{DateTopologicalInterpretationOf} are ongoing debates.  
Here we will consider the simplest possibility, namely the standard coupling of Dirac fermions
to vielbein gravity.

In $3+1$ dimensions the quantisation of this theory was carried out for the first time in the 
context of the new variables in \cite{ThiemannQSD5, ThiemannKinematicalHilbertSpaces}. The new ingredient was the passage 
to Gra{\ss}mann valued half densities and a representation in terms of holomorphic wave functions
of the fermionic variables. Technically, in $3+1$ dimensions one works in the time gauge 
and with the Ashtekar -- Barbero connection which can be obtained by an extension of the
ADM phase space subject to an SO$(3)$ Gau{\ss} constraint.

In higher dimensions, an Ashtekar -- Barbero like connection is not available and therefore 
a new idea is needed in order to arrive at a connection formulation with {\it compact gauge group}
although we are considering {\it Lorentzian} gravity. We start from the usual Dirac -- Palatini Lagrangian for Lorentzian General Relativity
and introduce the time gauge. This results in a formulation in terms of a canonical pair $(K_a^i,E^a_i),\;\;a,b,c,..=1,..,D;\;i,j,k,..=1,..,D$
which is subject to an SO$(D)$ Gau{\ss} constraint. 
We now extend this phase space by a canonical pair $(A_{aIJ},\pi^{aIJ})$ subject to 
the simplicity and SO$(D+1)$ Gau{\ss} constraint. This way we arrive at a connection formulation 
in terms of the compact gauge group SO$(D+1)$ although we are considering Lorentzian gravity.
Of course, the fermionic contribution to the Hamiltonian constraint of Lorentzian gravity,
just as in $3+1$ dimensions, acquires correction terms as compared to its Euclidean counterpart
which in part is due to switching from Lorentzian to Euclidean $\gamma$ matrices. 
Yet, these corrections are not as cumbersome as one might expect because the Lorentzian 
Clifford algebra differs from the Euclidean one just by a factor of $i$ in front of $\gamma^0$. 

After having obtained the fermionic contributions to the classical constraints we quantise them 
using standard methods \cite{ThiemannQSD5} and using the representation \cite{ThiemannKinematicalHilbertSpaces}.
Our formulation provides access to the  quantisation techniques developed in \cite{AshtekarRepresentationTheoryOf, AshtekarDifferentialGeometryOn, AshtekarProjectiveTechniquesAnd, MarolfOnTheSupport, AshtekarQuantizationOfDiffeomorphism} and allows for a rigorous construction of a quantum theory. Ideas on how to quantise the simplicity constraint, which has been shown to be anomalous in \cite{WielandComplexAshtekarVariables,BTTIII} are supplied in \cite{BTTV}.

\section{Canonical Analysis of Lorentzian Gravity coupled to Dirac Fermions}

As opposed to pure gravity where, in the end, it does not matter whether one starts with a first or a second order formulation of the theory, this choice results in inequivalent theories when dealing with fermions. The reason for this is that the torsion freeness condition which one derives when starting with first order general relativity, as e.g. done in our companion paper \cite{BTTII}, is modified by a term quadratic in the fermions, thus resulting in a non-vanishing torsion. At the end of the canonical analysis, one arrives at the same set of variables, but, after solving the equations of motion for the torsion part of the connection, one obtains more interaction terms, most prominently four-fermion interactions, which are not present in the theory when starting with a second order formulation. To the best of the authors' knowledge, it is unclear which type of action should be preferred on physical grounds. The second order variant leads to less interaction terms and could thus be preferred by demanding simplicity. On the other hand, when deriving the Ashtekar-Barbero variables from the Holst action, one deals in a first order framework and one could thus consider it more natural to choose this route. In this paper, we will choose the first order approach since the results of the canonical analysis of our companion paper \cite{BTTII} can be nicely used in order to deal with the torsion terms. For further literature on this topic, we refer to \cite{AshtekarInclusionOfMatter, MercuriFermionsInThe, ThiemannQSD5}.

We start with the first order action
\be S_{\text{G + F}} = -\int_\mathcal{M} d^{D+1} x \left( \frac{1}{2} e e^{\mu I} e^{\nu J} F_{\mu \nu IJ}(A) + \frac{i}{2} \overline{\Psi} e^\mu_I \gamma^I D_\mu \Psi - \frac{i}{2} \overline{ D_\mu \Psi} e^\mu_I \gamma^I \Psi \right) \text{.} \label{action} \ee
$\mathcal{M}$ denotes the space-time manifold of topology $\mathbb{R} \times \sigma$. $\mathcal{M}$ foliates into hypersurfaces $\Sigma_t := X_t(\sigma)$, where $X_t : \sigma \rightarrow M$ is an embedding defined by $X_t(\sigma) = X(t, \sigma)$. $x^a\text{,} \, a,b,c = 1, \ldots, D$ are local coordinates on $\sigma$. The signature of the space-time metric is given by $(-, +, \ldots, +)$. $e^{\mu I}$ denotes the vielbein and $F_{\mu \nu IJ} := \partial_\mu A_{\nu IJ} - \partial_\nu A_{\mu IJ} + [A_\mu, A_\nu]_{IJ}$ is the field strength of the SO$(1,D)$ connection $A_{\mu IJ}$. $\Psi$ denotes a Gra{\ss}mann valued Dirac spinor. Additionally, $D_\mu \Psi = \partial_\mu \Psi + \frac{i}{2} A_{aIJ} \Sigma^{IJ} \Psi$, $\Sigma^{IJ} = -\frac{i}{4}[\gamma^I, \gamma^J]$ and $\overline{\Psi} = \Psi^\dagger \gamma^0$. The properties of the $\gamma$ matrices are summarised in appendix \ref{appendix_gamma}. The gravitational part of this action has been analysed in \cite{BTTI,BTTII}, we will therefore concentrate on the fermionic part.

The split in space and time is performed analogously to the $D=3$ case. We split the time-evolution vector field into
\be  T^\mu = N n^\mu + N^\mu \text{,} ~~ n_\mu N^\mu = 0 \text{.} \ee 
$N$ is called the lapse function and $N^\mu$ the shift vector field. $n^\mu$ is the unit future pointing vector field normal to the spatial slices $\Sigma_t$, i.e. $n^\mu n_\mu = -1$ and $n^\mu \partial_\mu X_t = 0$. We use
\be  \delta^\mu \m_\nu= (g^\mu \m_\nu + n^\mu n_\nu) - n^\mu n_\nu =:q^\mu \m_\nu -  n^\mu n_\nu \ee 
to project the vielbein as
\be  e^\mu_I =q^\mu \m_\nu  e^\nu_I  - e^\nu_I n_\nu n^\mu =: \m^\| e^\mu_I - n_I n^\mu \text{.}\ee 

We choose the time gauge prior to the canonical analysis
(the time gauge is a canonical gauge, see, for instance, \cite{HenneauxQuantizationOfGauge}) by setting $n^I = \delta^I_0$. The action can now be rewritten as\footnote{A proper canonical analysis would 
work in terms of the momentum $\pi^{aIJ}$ conjugate to $A_{aIJ}$. However, here we use 
a shortcut and use the fact that the corresponding simplicity constraint \cite{BTTI,BTTII} imposes 
$\pi^{aIJ}=2 n^{I[} E^{aJ]} ,\;n^J n_J=-1,\;n_I E^{aI}=0$.} 
\be  S_{\text{G + F}} =  \int dt \int_\sigma d^{D} x \left( \dot{E}^{a}_i K_a^i + i(\sqrt[4]{q} \Psi^\dagger) (\sqrt[4]{q}\Psi) \dot{\m} - N \mathcal{H} - N^{a} \mathcal{H}_a - \lambda_{ij} G^{ij} - (\lambda_i + *_i) G^i \right) \text{,}  \ee 
where
\begin{eqnarray} 
  \mathcal{H} & = & \frac{1}{2} \sqrt{q} R + \frac{1}{\sqrt{q}} E^{[a|i} E^{b]j} K_{ai} K_{bj} + \frac{1}{\sqrt{q}} \frac{1}{8} \bar{K}^{\text{tr.fr.}}_{aij} F^{aij,bkl} \bar{K}^{\text{tr.fr.}}_{bkl} \nonumber \\ \nonumber &  & + \frac{i}{2} \frac{1}{\sqrt{q}} \overline{\sqrt[4]{q} \Psi} E^a_i \gamma^i \nabla_a (\sqrt[4]{q}\Psi) - \frac{i}{2}    \frac{1}{\sqrt{q}}  \overline{ \nabla_a (\sqrt[4]{q}\Psi)} E^a_i \gamma^i (\sqrt[4]{q}\Psi) \\
  & & - \frac{1}{\sqrt{q}} (\sqrt[4]{q}\Psi)^\dagger \Sigma^{ij} (\sqrt[4]{q}\Psi) K_{ai} E^{a}_j - \frac{1}{4} \frac{1}{\sqrt{q}} \overline{\sqrt[4]{q} \Psi} E^a_k  \left\{ \gamma^k , \Sigma^{ij} \right\} (\sqrt[4]{q}\Psi) \bar{K}^{\text{tr.fr.}}_{aij}  \text{,} \\
  \mathcal{H}_a & = & -2 E^{bj} \nabla_{[a} K_{b]j}+\frac{i}{2} (\sqrt[4]{q}\Psi)^\dagger \nabla_a (\sqrt[4]{q}\Psi) -\frac{i}{2} (\nabla_a (\sqrt[4]{q}\Psi))^\dagger (\sqrt[4]{q}\Psi) + \frac{1}{2} \bar{K}^{\text{tr.fr.}}_{aij} G^{ij}  \text{,} \\
  G^{ij} & = & 2 K_{a}^{[i} E^{a|j]} - (\sqrt[4]{q}\Psi)^\dagger \Sigma^{ij} (\sqrt[4]{q}\Psi) \text{,} \\
  G^{i}  & = & \bar{K}^{\text{tr.}}_{aij} E^{aj} \text{,}
\end{eqnarray}
and small Latin indices $i,j,k,\ldots = 1,\ldots,D$ are internal indices in the time gauge. $F^{aij, bkl}$ and the derivation of the symplectic structure have been described in $\cite{BTTI,BTTII}$. The Gau{\ss} constraint has been split into its rotational part $G^{ij}$ and its boost part $G^{i}$. $\lambda_{ij} := - T^\mu A_{\mu ij} $ and $\lambda_i = \lambda_{i0}$. All terms proportional to $\bar{K}^{\text{tr.}}_{aij} E^{aj}$ not belonging to the boost part of the Gau{\ss} constraint have been written as $*_i G^i$. Further, $E^{ai} = \sqrt{q} \m^\| e^{a i}$ and we have decomposed $A_{aIJ} = \Gamma_{aIJ} + 2 n_{[I} K_{a|J]} + \bar{K}_{aij}$. The bar notation $\bar{K}_{aij}$ which was already used in our companions papers \cite{BTTI, BTTII} means that the internal indices are orthogonal on $n^I$. In the time gauge, this is equivalent of having only small latin indices running from $1$ to $D$. The splits in trace and trace free parts are done with respect to the vielbein, e.g. $\bar{K}^{\text{tr.}}_{aij} = \frac{1}{D} \bar{K}^{\text{tr.}}_{bkj} E^{bk} E_{ai}$. The boost part of the Gau{\ss} constraint does not acquire a fermionic part because of the cancellation $\Sigma^{0i} + (\Sigma^{0i})^\dagger = 0$. The Dirac spinors in the above equations appear only as half-densities, i.e. $\sqrt[4]{q} \Psi$. Since the symplectic structure tells us that these half-densities are the natural canonical variables, we will abuse notation and denote by $\Psi$ from now on a half-density. The importance of using half densities stems from the simple form of the symplectic structure. Otherwise, the connection would acquire a complex part \cite{JacobsonFermionsInCanonical} and the techniques introduced in \cite{AshtekarRepresentationTheoryOf, AshtekarDifferentialGeometryOn, AshtekarProjectiveTechniquesAnd, MarolfOnTheSupport, AshtekarQuantizationOfDiffeomorphism} would not be accessible. 

In order to facilitate the canonical analysis, we will employ the equations of motion for $\lambda_i$ and $\bar{K}^{\text{tr.fr.}}_{aij}$ at the Lagrangian level. Their solutions translate directly to a purely canonical treatment as one can check. Variation of the Lagrangian with respect to $\lambda_i$ sets the boost part of the Gau{\ss} constraint to zero. Variation with respect to $\bar{K}^{\text{tr.fr.}}_{aij}$ yields
\be  \bar{K}^{\text{tr.fr.}}_{aij} = F^{-1}_{aij, bkl} \overline{\Psi} E^b_m  \left\{ \gamma^m , \Sigma^{ij} \right\} \Psi  \text{,} \ee 
which we use to eliminate $\bar{K}^{\text{tr.fr.}}_{aij}$ in $\mathcal{H}$.

Next, we perform the Legendre transform, yielding the constraints 
\begin{eqnarray} 
  \mathcal{H} & = & \frac{1}{2} \sqrt{q} R + \frac{1}{\sqrt{q}} E^{[a|i} E^{b]j} K_{ai} K_{bj} \nonumber  \\ \nonumber &  & + \frac{i}{2} \frac{1}{\sqrt{q}} \overline{\Psi} E^a_i \gamma^i \nabla_a \Psi - \frac{i}{2}    \frac{1}{\sqrt{q}}  \overline{ \nabla_a \Psi} E^a_i \gamma^i \Psi \\
  & & - \frac{1}{2} \Psi^\dagger \Sigma^{ij} \Psi\Psi^\dagger \Sigma_{ij} \Psi + \frac{1}{32} \overline{\Psi}   \left\{ \gamma^k , \Sigma^{ij} \right\} \Psi \overline{\Psi}   \left\{ \gamma_k , \Sigma_{ij} \right\} \Psi  \text{,} \\ 
  \mathcal{H}_a & = & -2 E^{bj} \nabla_{[a} K_{b]j}+\frac{i}{2} \Psi^\dagger \nabla_a \Psi -\frac{i}{2} (\nabla_a \Psi)^\dagger \Psi  \text{,} \\
  G^{ij} & = & 2 K_{a}^{[i} E^{a|j]} - \Psi^\dagger \Sigma^{ij} \Psi \text{,} 
\end{eqnarray}
where $\nabla_a$ is the covariant derivative associated to the spin connection $\Gamma_{aij}$, as well as the non-vanishing (generalised) Poisson (anti-) brackets \cite{HenneauxQuantizationOfGauge}
\be  \{ E^{ai}(x), K_{bj}(y)\} = \delta^D(x-y) \delta^a_b \delta^i_j  ~~~ \text{and} ~~~ \{ \Psi^\alpha(x), -i \Psi^\dagger_\beta(y) \} = - \delta^D(x-y) \delta^\alpha_\beta \text{.}  \ee 
A term proportional to the Gau{\ss} constraint has been omitted in $\mathcal{H}$ and $\mathcal{H}_a$. 

We define the generator of spatial diffeomorphisms 
\be  \tilde{\mathcal{H}}_a := \mathcal{H}_a - \frac{1}{2} \Gamma_{aij} G^{ij} = - E^{bj} \partial_a K_{bj} + \partial_b (E^{bj} K_{aj} ) + \frac{i}{2} \Psi^\dagger \partial_a \Psi -\frac{i}{2} (\partial_a \Psi^\dagger ) \Psi  \text{,} \ee 
which acts as 
\begin{eqnarray}
\left\{ E^{ai}, \tilde{\mathcal{H}}_b[N^b] \right\} & = & N^b \partial_b E^{ai} + (\partial_b N^b) E^{ai}- (\partial_b N^a) E^{bi} \text{,}\\
\left\{ K_{ai}, \tilde{\mathcal{H}}_b[N^b] \right\} & = & N^b \partial_b K_{ai} + (\partial_a N^b) K_{bi}\text{,}\\
\left\{ \Psi, \tilde{\mathcal{H}}_b[N^b] \right\} & = & N^b \partial_b \Psi + \frac{1}{2} (\partial_a N^a) \Psi \text{,}\\
\left\{ \Psi^\dagger, \tilde{\mathcal{H}}_b[N^b] \right\} & = & N^b \partial_b \Psi^\dagger + \frac{1}{2} (\partial_a N^a) \Psi^\dagger \text{,} \\
\end{eqnarray}
by Lie derivatives. The Gau{\ss} constraint acts as
\begin{eqnarray} 
\left\{ E^{ai}, \frac{1}{2} G^{ij}[\lambda_{ij}] \right\} & = & \lambda^i \m_j E^{aj} \text{,}\\ 
\left\{ K_{ai}, \frac{1}{2} G^{ij}[\lambda_{ij}] \right\} & = & \lambda_i \m^j K_{aj} \text{,} \\ 
\left\{ \Psi, \frac{1}{2} G^{ij}[\lambda_{ij}] \right\} & = & \frac{1}{2} i \lambda_{ij} \Sigma^{ij} \Psi \text{,} \\ 
\left\{ \Psi^\dagger, \frac{1}{2} G^{ij}[\lambda_{ij}] \right\} & = & -\frac{1}{2} i \Psi^\dagger \lambda_{ij} \Sigma^{ij} \text{,} \\
\left\{ D_a \Psi, \frac{1}{2} G^{ij}[\lambda_{ij}] \right\} & = & \frac{1}{2} i \lambda_{ij} \Sigma^{ij} D_a \Psi \text{,} \\ 
\left\{ (D_a \Psi)^\dagger, \frac{1}{2} G^{ij}[\lambda_{ij}] \right\} & = & -\frac{1}{2} i (D_a \Psi)^\dagger \lambda_{ij} \Sigma^{ij} \text{,}  \\
\left\{ \Psi^\dagger \Sigma^{ij} \Psi, \frac{1}{2} G^{ij}[\lambda_{ij}] \right\} & = & \Psi^\dagger [\lambda, \Sigma]^{ij} \Psi \text{,} \\  
\left\{ \overline{\Psi} \{ \gamma^k, \Sigma^{ij} \} \Psi, \frac{1}{2} G^{ij}[\lambda_{ij}] \right\} & = & \overline{\Psi} \left( \{ \gamma^k, [\lambda ,\Sigma]^{ij} \} + \{\lambda_{km} \gamma^m, \Sigma^{ij} \} \right) \Psi  \text{.} 
\end{eqnarray}

We therefore conclude that the algebra of the diffeomorphism and Gau{\ss} constraints closes and that they both Poisson-commute with the Hamiltonian constraint, at least weakly.

Thus we are left with checking the Poisson bracket of two Hamiltonian constraints. We split $\mathcal{H} = \mathcal{H}_{\text{grav}} + \mathcal{H}_{\text{2F}} + \mathcal{H}_{\text{4F}}$ into the purely gravitational part, a part containing two fermions and a part containing the four-fermion terms and define $V_a := M \partial_a N - N \partial_a M$ as well as $V_{ab} := (\partial_a M) (\partial_b N) - (\partial_b M) (\partial_a N)$. The non-vanishing Poisson brackets are given as
\begin{eqnarray}
 \left\{ \mathcal{H}_{\text{grav}}[M], \mathcal{H}_{\text{grav}}[N] \right\} &=& \int_\sigma  d^Dx \Biggl( V_a q^{ab} \left(-2 E^{cj} \nabla_{[b} K_{c]j} \right)    \\ & & ~~~~~~~~~~~ + V_{ab} \frac{E^{ai} E^{bj}}{q} K_{a[i} E^a_{j]} \Biggr) \nonumber  \text{,} \\
 \left\{ \mathcal{H}_{\text{2F}}[M], \mathcal{H}_{\text{2F}}[N] \right\} &=& \int_\sigma  d^Dx \Biggl( V_a q^{ab} \left( \frac{i}{2} \Psi^\dagger \nabla_a \Psi -\frac{i}{2} (\nabla_a \Psi)^\dagger \Psi \right) ~~~   \nonumber \\ & & ~~~~~~~~~~~ - V_{ab} \frac{E^{ai} E^{bj}}{2q} \Psi^\dagger \Sigma^{ij} \Psi \Biggr) \nonumber \text{,} \\
  \left\{ \mathcal{H}_{\text{grav}}[M], \mathcal{H}_{\text{2F}}[N] \right\} + \left\{  \mathcal{H}_{\text{2F}}[M], \mathcal{H}_{\text{grav}}[N] \right\}&=& \int_\sigma  d^Dx \left(\frac{1}{8q} V_a \overline{\Psi} \{ E^a_k \gamma^k, \Sigma^{ij} \} \Psi \Psi^\dagger \Sigma_{ij} \Psi \right)  \nonumber \text{,}\\
    \left\{ \mathcal{H}_{\text{2F}}[M], \mathcal{H}_{\text{4F}}[N] \right\} + \left\{ \mathcal{H}_{\text{4F}}[M], \mathcal{H}_{\text{2F}}[N] \right\} &=& \int_\sigma  d^Dx \left(-\frac{1}{8q} V_a \overline{\Psi} \{ E^a_k \gamma^k, \Sigma^{ij} \} \Psi \Psi^\dagger \Sigma_{ij} \Psi \right)  \nonumber \text{,}
\end{eqnarray}
and sum up to 
 \be  \left\{ \mathcal{H}[M], \mathcal{H}[N] \right\} = \int_\sigma  d^Dx \left( V_a q^{ab} \mathcal{H}_b + V_{ab} \frac{E^{a}_i E^{b}_j}{2q}G^{ij}\right) \text{.}  \ee 
The constraints are therefore consistent and the canonical analysis ends here.

\section{Phase Space Extension}

In \cite{BTTI,BTTII} it was shown that the extension of the ADM phase space $(q_{ab}, P^{ab})$ to the extended phase space $(A_{aIJ}, \pi^{aIJ})$ subject to  Gau{\ss} and simplicity constraint
is equivalent to the ADM phase space. Moreover, this is possible using SO$(D+1)$ as the structure group while considering Lorentzian gravity. Since spinors can only be coupled to vielbeins, we have to construct a transformation from $(E^{ai}, K_{ai})$ to $(A_{aIJ}, \pi^{aIJ})$. The calculation turns out to be very similar to the one described in \cite{BTTI,BTTII}, we therefore only give the result and comment on some peculiarities. 

The explicit construction is given by
\be  \bar{E}^{aI} = \zeta \bar{\eta}^{I} \m_J \pi^{aKJ} n_K, ~~~ \bar{K}_{aI} = \zeta \bar{\eta}_I \m^J (A-\Gamma)_{aKJ} n^K \text{,}  \ee 
where $\bar{\eta}^{IJ} = \eta^{IJ}-\zeta n^I n^J \approx \eta^{IJ} - \frac{\zeta}{D-1} \left( \pi^{aKI} \pi_{aK} \m^J -  \zeta \eta^{IJ} \right)$, $\Gamma_{aIJ}$ is the hybrid connection of 
$E^{aI}$ (see \cite{BTTI,BTTII} for details) and  $ \eta^{00} = \zeta$, $\eta^{ij} = \delta^{ij}$. The peculiarity of these expressions is the appearance of $n^I$, which can only 
be directly (that is, without non-polynomial terms except for $\sqrt{q}$) expressed in terms of  $\pi^{aIJ}$ for $D+1$ odd. For general $D$, we only have access to $n^I n^J$ and then can define 
$\pm n^I$ through $\pm n^I = \sqrt{n^I n^I} \text{sgn}(n^0 n^I)$ (no summation understood here and one substitutes for 
$n^I n^I$ under the square root the expression for $n^I n^J$ at $I=J$). Fortunately, we 
can avoid to make use of this explicit square root expression  by invoking the following trick: Ultimately the non-vanishing Poisson bracket involving $n^I$ is of the form $n^J \{A_{aIJ},n^K\}$. Since 
$n^K n_K\approx \zeta$ modulo simplicity constraint we have $n^J \{A_{aIJ}, n^K\} n_K\approx 0$.
To see this, notice that the simplicity constraint reads $S^{cd}_{\overline{M}}=\epsilon_{IJKL\overline{M}} \pi^{cIJ} \pi^{dKL}$ (see \cite{BTTI,BTTII} for details). It follows 
$$
n^J \{A_{aIJ},S^{cd}_{\overline{M}}\}=2n^J \epsilon_{IJKL\overline{M}} \delta_a^{(c} 
\pi^{d)KL}\approx 0
$$
on the constraint surface $\pi^{aIJ}=2 n^{[I} E^{aJ]} $. It follows 
$n^J \{A_{aIJ},n^K\}\approx n^J \{A_{aIJ},n^L\} \bar{\eta}^K_L=- n^J \{A_{aIJ},\bar{\eta}^K_L\} n^L$. 
However, $\{A_{aIJ},\bar{\eta}^K_L\}=-\zeta\{A_{aIJ}, n^K n_L\}$ and $n^K n_L$ can be 
expressed unambiguously as above in terms of $\pi^{aIJ}$. In order to compute the brackets between $\bar{E}^{aI},
\bar{K}_{aI}$ one then just hast to carefully insert the definition of $n_I n_J$ in terms of $\pi^{aIJ}$.
The only term which cannot easily be seen to vanish by algebraic manipulations alone occurs in 
the bracket $\{K_{aI}, K_{aJ}\}$ and is of the form
$$
\bar{\eta}_I^K n^L \bar{\eta}_J^M n^N \{[A-\Gamma]_{aIJ},[A-\Gamma]_{bKL}\}
=-\bar{\eta}_I^K n^L \bar{\eta}_J^M n^N 
[\{[A_{aIJ},\Gamma_{bKL}\}-\{A_{bKL},\Gamma_{aIJ}\}] \text{.}
$$
This term vanishes due to the weak integrability (modulo simplicity constraint) of the 
hybrid connection $\Gamma_{aIJ}$ and by using the trick mentioned above. See \cite{BTTI}
for more details.\\
\\
After a tedious calculation, 
the Poisson brackets of $\bar{E}^{aI}$ and $\bar{K}_{aI}$ expressed as functions of $A_{aIJ}$ and $\pi^{aIJ}$ are given by
\be  \{ \bar{E}^{aI}(x),  \bar{E}^{bJ}(y)\} = 0, ~~~ \{ \bar{K}_{aI}(x),  \bar{K}_{bJ}(y) \} \approx 0,  ~~~ \{ \bar{E}^{aI}(x), \bar{K}_{bJ}(y) \} \approx -\zeta \delta^D(x-y) \delta^a_b \bar{\delta}^I_J \text{.}  \ee 
modulo simplicity constraint.\\
\\
The only task left to do is to write down a Hamiltonian theory in the variables $A_{aIJ}$ and $\pi^{bKL}$ with internal gauge group SO$(D+1)$ which reduces to the theory derived in the previous section on the constraint surface $S^{ab}_{\overline{M}} = n_I G^{IJ} = 0$. The basic idea is to first derive a Hamiltonian formulation of Euclidean gravity coupled to fermions and then to adjust the Hamiltonian constraint to mimic Lorentzian gravity. The reason why the procedure introduced in \cite{BTTI,BTTII} generalises nicely to Dirac fermions is the strong resemblance of the Clifford algebras, which differ only by factors of $i$ for different signatures and the Euclidean signature of the internal gauge group which ensures that $\Sigma^{IJ}$ is a Hermitian matrix the Euclidean case. 
This requires care at several places, e.g. the cancellation $\Sigma^{0i} + (\Sigma^{0i})^\dagger = 0$ from the boost part of the Lorentzian Gau{\ss} constraint is no longer present. In order to derive the Euclidean constraints, we start as in the previous section with the action (\ref{action}) and perform a $D+1$ decomposition. We replace $\gamma^0$ with $n_I \gamma^I$, which reduces to $\gamma^0$ in the time gauge. We note that the object $\overline{\Psi} \Psi$ is not a Lorentz scalar any more when using Euclidean signature, because the $\gamma^0$ inherent in 
$\overline{\Psi}\Psi$ is needed in order to maintain invariance under boosts which are 
generated by the anti-Hermitian $\Sigma^{0i}$. In Euclidean signature the boost generator is also
Hermitian and thus $\Psi^\dagger \Psi$ rather than $\overline{\Psi}\Psi$ is now the appropriate Euclidean scalar to be used while $\Psi^\dagger \gamma^I \Psi$ is a Euclidean covariant vector with index $I$. The substitution $\gamma^0 \rightarrow n_I \gamma^I$ is therefore natural for Euclidean signature and allows for the construction of a manifestly SO$(D+1)$ gauge invariant theory. We use the additional $n^I$ in the action to form $\pi' \m^{aIJ} = 2 n^{[I} E^{a|J]}$ (see \cite{BTTI,BTTII} for details) and introduce the simplicity constraint in order to replace $\pi^{\prime aIJ}$ by  
$\pi^{aIJ}$. The Euclidean Hamiltonian theory is then given by the constraints
\begin{eqnarray}
 \mathcal{H}^E & = & \frac{1}{2} \pi^{aIK} \pi^{bJ} \m_K F_{abIJ} + \left( \frac{1}{2} \Psi^\dagger \pi^{aIJ} \Sigma_{IJ} D_a \Psi + CC \right) \text{,}\\
 \mathcal{H}^E_a & = & \frac{1}{2} \pi^{bIJ} F_{abIJ} + \frac{i}{2} \Psi^\dagger D_a \Psi - \frac{i}{2} (D_a \Psi)^\dagger \Psi \text{,} \\
 G_E^{IJ} & = & D_a \pi^{aIJ}- \Psi^\dagger \Sigma^{IJ} \Psi \text{,} \\
 S^{ab}_{\overline{M}} &=& \frac{1}{4} \epsilon_{IJKL\overline{M}} \pi^{aIJ} \pi^{bKL} \text{,} 
\end{eqnarray}
and the Poisson bracket 
\be  \{ A_{aIJ}(x), \pi^{bKL}(y) \} = \delta^D(x-y) \delta^b_a (\delta^I_K \delta^J_L - \delta^I_L \delta^J_K), ~~~~ \{ \Psi^\alpha(x), -i \Psi^\dagger_\beta(y) \} = - \delta^D(x-y) \delta^\alpha_\beta \text{.}  \ee 
The task of ``Lorentzifying'' the gravitational part of $\mathcal{H}^E$ has been addressed in \cite{BTTI,BTTII}. For the fermionic part, we observe that we should add a factor of $i$ in front of the fermionic term in order to compensate for $\gamma^0_E = i \gamma^0_L$ and denote the changed constraint by $\mathcal{H}^E_{(i)}$. The Hamiltonian constraint now reduces to 
\begin{eqnarray}
 \mathcal{H}^E_{(i)} & = & \frac{1}{2} \sqrt{q} R -  \frac{1}{\sqrt{q}} E^{[a|i} E^{b]j} K_{ai} K_{bj} + \left( \frac{i}{2\sqrt{q}} \Psi^\dagger \gamma_L^0 E^{a}_i \gamma^i \nabla_a \Psi + CC \right) \nonumber \\
  & & - \frac{1}{2\sqrt{q}} \Psi^\dagger \Psi E^{ai} K_{ai} - \frac{1}{2 \sqrt{q}} \frac{D-2}{D-1}\Psi^\dagger \Sigma^{0i}_E \Psi \Psi^\dagger \Sigma_{0i}^E \Psi + \partial_a \left( \frac{E^a_i}{\sqrt{q}} \Psi^\dagger \Sigma_E^{0i} \Psi \right) + \mathcal{O}(K^{\text{tr.fr.}}_{aij}) \text{.} ~~~~~~ 
\end{eqnarray}
The terms proportional to $K^{\text{tr.fr.}}_{aij}$ can be dealt with using ideas from gauge unfixing. We calculate
\begin{eqnarray}
 \{ S^{ab}_{\overline{M}}[c_{ab}^{\overline{M}}], \mathcal{H}^E_{(i)}[N] \} &=& D^{ab}_{\overline{M}}[Nc_{ab}^{\overline{M}}] + \int_\sigma d^D x \frac{N}{4} c_{ab}^{\overline{M}} \pi^{aIJ} \pi^{bMN} \epsilon_{MNKL\overline{M}} \Psi^\dagger \{ \Sigma_{IJ}, \Sigma_{KL} \} \Psi  \nonumber \\ &:=& D^{ab}_{\overline{M}}[Nc_{ab}^{\overline{M}}] + \m_F D^{ab}_{\overline{M}}[Nc_{ab}^{\overline{M}}]
\end{eqnarray}
and see that the gravitational constraint $ D^{ab}_{\overline{M}} =  -\epsilon_{IJKL\overline{M}} \pi^{cIJ} \left( \pi^{(a|KN} D_c \pi^{b)L}\m_N \right) \approx 0$ now receives a fermionic contribution $\m_F D$. $F^{ab}_{\overline{M}} \m^{cd}_{\overline{N}}$, the Dirac matrix introduced in \cite{BTTII}, however remains unchanged since $\m_F D$ Poisson-commutes with the simplicity constraint and gauge unfixing works as before. Next to compensating the terms proportional to $K^{\text{tr.fr.}}_{aij}$, gauge unfixing also produces a four-fermion term, which we have to subtract again in order 
to build the correct Lorentzian Hamiltonian constraint. 

Comparison with the previous section leads to the following correction terms:
\begin{eqnarray}
 \mathcal{H}^L & = & \mathcal{H}^E_{(i)} +\frac{2}{\sqrt{q}} E^{[a|I} E^{b]J} K_{aI} K_{bJ} - \frac{1}{2} \m_G D^{ab}_{\overline{M}} \; \left( F^{-1} \right) \m_{ab}^{\overline{M}} \, \m_{cd}^{\overline{N}} \; \m_G D^{cd}_{\overline{N}} \nonumber \\
 & & - \frac{1}{2} \m_G D^{ab}_{\overline{M}} \; \left( F^{-1} \right) \m_{ab}^{\overline{M}} \, \m_{cd}^{\overline{N}} \; \m_F D^{cd}_{\overline{N}} - \frac{1}{2} \m_F D^{ab}_{\overline{M}} \; \left( F^{-1} \right) \m_{ab}^{\overline{M}} \, \m_{cd}^{\overline{N}} \; \m_G D^{cd}_{\overline{N}} \nonumber \\
 & & - \frac{1}{2 \sqrt{q}} \Psi^\dagger \Sigma^{ij} \Psi \Psi^\dagger \Sigma_{ij} \Psi + \frac{1}{2 \sqrt{q}} \frac{D-2}{D-1}\Psi^\dagger \Sigma^{0i}_E \Psi \Psi^\dagger \Sigma_{0i}^E \Psi \nonumber \\
  & & - \partial_a \left( \frac{E^a_i}{\sqrt{q}} \Psi^\dagger \Sigma_E^{0i} \Psi \right) + \frac{1}{2\sqrt{q}} \Psi^\dagger \Psi E^{aI} K_{aI} + \frac{1}{32} \overline{\Psi}   \big\{ \gamma^k , \Sigma^{ij} \big\} \Psi \overline{\Psi}   \big\{ \gamma_k , \Sigma_{ij} \big\} \Psi  \text{.}  
\end{eqnarray}

This Hamiltonian has to be rewritten in terms of $A_{aIJ}$ and $\pi^{bKL}$ only, desirably as simple as possible regarding the quantisation. We propose the Hamiltonian

\begin{eqnarray}
 \mathcal{H}^L & = & \frac{1}{2} \pi^{aIK} \pi^{bJ} \m_K F_{abIJ} + \left( i \frac{1}{2} \Psi^\dagger \pi^{aIJ} \Sigma_{IJ} D_a \Psi + CC \right) \nonumber  \\ 
 & & +\frac{2}{\sqrt{q}} E^{[a|I} E^{b]J} K_{aI} K_{bJ} - \frac{1}{2} \m_G D^{ab}_{\overline{M}} \; \left( F^{-1} \right) \m_{ab}^{\overline{M}} \, \m_{cd}^{\overline{N}} \; \m_G D^{cd}_{\overline{N}}  \nonumber \\
 & & - \frac{1}{2} \m_G D^{ab}_{\overline{M}} \; \left( F^{-1} \right) \m_{ab}^{\overline{M}} \, \m_{cd}^{\overline{N}} \; \m_F D^{cd}_{\overline{N}} - \frac{1}{2} \m_F D^{ab}_{\overline{M}} \; \left( F^{-1} \right) \m_{ab}^{\overline{M}} \, \m_{cd}^{\overline{N}} \; \m_G D^{cd}_{\overline{N}} \nonumber \\
 & & - \frac{1}{2 \sqrt{q}} \Psi^\dagger \Sigma^{IJ} \Psi \Psi^\dagger \Sigma_{IJ} \Psi + \frac{1}{2 \sqrt{q}} \frac{3D-4}{D-1}\Psi^\dagger \Sigma^{IK} \Psi \Psi^\dagger \Sigma_{JK} \Psi n_I n^J \\
  & & - \partial_a \left( \frac{\pi^{a} \m_{IJ}}{\sqrt{q}} \Psi^\dagger \Sigma^{IJ} \Psi \right) + \frac{1}{2\sqrt{q}} \Psi^\dagger \Psi E^{ai} K_{ai} - \frac{1}{32} \Psi^\dagger  \gamma^{[I} \gamma^J \gamma^K \gamma^{L]} \Psi n_L n^M \Psi^\dagger  \gamma_{[I} \gamma_J \gamma_K \gamma_{M]} \Psi    \nonumber 
\end{eqnarray}
for quantisation, although we are well aware of the fact that other choices might lead to equally reasonable classical starting points. We shift the problem of choosing the ``correct'' Hamiltonian constraint to the semiclassical analysis. The expressions for $n^I n_J$ and $E^{aI} K_{bI}$ were derived in \cite{BTTI,BTTII} and all $\gamma$-matrices appearing are those for Euclidean signature.

\section{Kinematical Hilbert Space for Fermions}

The construction of the kinematical Hilbert space for fermions was discussed in \cite{ThiemannKinematicalHilbertSpaces}. Results obtained there apply for the case at hand, so we only give a short summary. It is crucial to work with half-densitised fermionic variables $\Psi$ for what follows, as was stressed in \cite{ThiemannKinematicalHilbertSpaces}. 

Faithful implementation of the reality conditions enforces the use of a representation in which the objects
\be \theta_{\alpha}(x) := \int_{\sigma} d^Dy \sqrt{\delta(x,y)} ~ \Psi_{\alpha}(y) := \lim_{\epsilon \rightarrow 0} \int_{\sigma}d^Dy \frac{\chi_{\epsilon}(x,y)}{\sqrt{\epsilon^D}}\Psi_{\alpha}(y)\ee
become densely defined multiplication operators. Their adjoints $\overline{\theta}^\alpha$ become derivative operators. Here, $\alpha = 1,...,n := 2^{\left\lfloor(D+1)/2\right\rfloor}$ ($\left\lfloor . \right \rfloor$ denotes the integer part of $.$) and $\chi_{\epsilon}(x,y)$ denotes the characteristic function of a box of Lebesgue measure $\epsilon^D$ centered at $x$. In the above equation, the half-densities $\Psi$ are ``dedensitised'' using the $\delta$-distribution, which is a scalar in one of its arguments and a density in the other. Thus, the variables $\theta$ are Gra{\ss}mann-valued scalar quantities, which is important for diffeomorphism invariance \cite{ThiemannKinematicalHilbertSpaces}. In calculations it is understood that the $\epsilon \rightarrow 0$ limit is performed after the manipulation under consideration is performed.

The variables $\theta_{\alpha}(v)$ coordinatise together with their conjugates the superspace $S_v$ at the point $v$. The quantum configuration space is the uncountable direct product $\overline{\mathcal{S}} := \prod_{v \in \sigma} S_v$. In order to define an inner product on $\overline{\mathcal{S}}$, it turns out to be sufficient to define an inner product on each $S_v$ coming from a probability measure. The ``measure'' on $S_v$ is a modified form of the Berezin symbolic integral \cite{Choquet-BruhatAnalysisManifoldsAnd2}

\be dm\left(\overline{\theta},\theta\right) = d\overline{\theta} d\theta e^{\overline{\theta}\theta} \text{ and } dm_v = \otimes_{\alpha=1}^{n}dm\left(\overline{\theta}_{\alpha}(v),\theta_{\alpha}(v)\right) \text{,}\ee
which has the additional property of being positive on holomorphic functions (those which only depend on the $\theta_{\alpha}$ and not on $\overline{\theta}_{\alpha}$). Since the $\theta$ are Gra{\ss}mann variables and thus anti-commute, any product of more than $n$ of these variables will vanish. The vector space of monomials of order $k$ is $n!/k!(n-k)!$-dimensional ($0 \leq k \leq n$) and the full vector space $Q_v$ built from all monomials has dimension $2^{n}$. The full fermionic Hilbert space is a space of holomorphic square integrable functions on $\overline{\mathcal{S}}$ with respect to $d\mu_F$
\be \mathcal{H}_F = L_2\left( \overline{\mathcal{S}}, d\mu_F \right) = \otimes_{v \in \sigma} L_2\left( S_v, dm_v\right)\text{.}\ee

When restricted to a point $v$, the inner product can be seen to coincide with the standard inner product on $Q_v$ when viewed as a vector space of exterior forms of maximal degree $D+1$. For a more complete treatment, the reader is referred to \cite{ThiemannKinematicalHilbertSpaces} where it is shown that the fermion measure $d\mu_F$ is gauge and diffeomorphism invariant and that the reality conditions $\overline{\theta}_{\alpha} = -i\pi_{\alpha}$ are faithfully implemented in the inner product.

\section{Implementation of the Hamiltonian Constraint Operator}

The quantisation of the purely gravitational Hamiltonian constraint in dimensions $D+1\geq 3$ was described in \cite{BTTIII} and we will not repeat it here. The quantisation of fermionic degrees of freedom was described in detail in \cite{ThiemannQSD5, ThiemannKinematicalHilbertSpaces}, which we assume the reader to be familiar with. Next to an explicit example, we will only provide a toolkit to quantise the fermionic part of the Hamiltonian constraint operator as writing down the explicit terms is rather laborious. 

Quantisation of the $\theta$ variables is performed by promoting $\theta_\alpha$ to a multiplication operator and $\widehat{(\theta^\dagger)^\beta} = -\hbar \frac{\partial^L}{\partial \theta_\beta}$, where $L$ indicates the left derivative. 
The explicit quantisation follows the (extended) toolkit of \cite{BTTIII}:
\renewcommand{\labelenumi}{(\arabic{enumi})}
\begin{enumerate}
  \item Choose a triangulation $T(\gamma, \epsilon)$ of the spatial slice $\sigma$ adapted to the graph $\gamma$. 
  \item Use classical identities in order to express the Hamiltonian constraint in terms of connections $A_{aIJ}$, volumes $V(x, \epsilon)$ and Euclidean Hamiltonian constraints $\mathcal{H}_E(x, \epsilon)$. 
  \item Replace all phase space variables by their corresponding regulated quantities.
  \item Instead of the the integration $\int_\sigma d^Dx$, put a sum $\frac{1}{D!} \sum_{v \in V(\gamma)}$ over all the vertices $v$ of the graph $\gamma$. 
  \item For every spatial $\epsilon$-symbol, put a sum $\frac{2^D}{E(v)} \sum_{v(\Delta)=v}$ over all $D$-simplices having $v$ as a vertex. The holonomies associated with the $\epsilon$-symbol are evaluated along the edges spanning $\Delta$.
    \item Substitute the generalised Poisson (anti-)brackets by $\frac{i}{\hbar}$ times the (anti-)commutator of the corresponding operators, i.e. the multiplication operator $\hat{h}_e$, the volume operator $\hat{V}$, the multiplication operator $\hat{\theta}_\alpha$ and the derivation operator $-\hbar \frac{\partial^L}{\partial \theta_\alpha}$. 
\end{enumerate}  

The kinetic fermionic part of the Hamiltonian constraint operator is a bit more involved since it contains a derivative. Following \cite{ThiemannKinematicalHilbertSpaces}, we explicitly get 
\begin{alignat}{3} 
  & \hat{H}^\epsilon_{\text{Dirac, kin}}(N) f_\gamma    \\
=&  \biggl( \frac{i \hbar}{2} \sum_{v \in V(\gamma)} \frac{2^D}{D!} \frac{N_v}{E(v)} \sum_{v(\Delta) = v} \m^{\hat{\epsilon}} \left( \frac{\pi^{a} \m_{IJ}}{\sqrt{q}} (v)\right) \left(\Sigma^{IJ} \left((h_{s_a(\Delta)} \theta(s_a(\Delta)(\epsilon) \right) -\theta(v)  \right)_\alpha \frac{\partial^L}{\partial \theta_\alpha(v)} +H.C. \biggr) f_\gamma \text{,} \nonumber 
\end{alignat}
where by $\m^{\hat{\epsilon}}(\ldots)$ we mean the regulated quantity with the Poisson brackets substituted by $i/\hbar$ times the commutator of the corresponding operators. The Hermitian conjugation operation $H.C.$ is meant with respect to the inner product on the Hilbert space. 

Due to its length, we refrain from writing down the complete Hamiltonian constraint operator which can be easily done when following the quantisation recipe. We remark that we could split the Dirac fermions for $D+1$ even into left- and right-handed parts, however, the presentation does not benefit from this. Details are supplied in \cite{BTTVI}.The quantisation ambiguities from loop quantum gravity are also present when considering fermions and, as usual, we shift this problem to the semiclassical limit.

\section{Outline of the Quantum Simplicity Constraints}

In this section, we briefly outline the implementation of the quantum simplicity constraint. As shown in \cite{BTTIII}, the quadratic simplicity constraint can be implemented on edges anomaly free and the result coincides with the spin foam treatment by Freidel, Krasnov and Puzio \cite{FreidelBFDescriptionOf}. The vertex constraints are second class when using holonomies and fluxes as basic variables and thus require a more elaborate treatment. One possibility is the master constraint approach discussed in \cite{BTTIII}. Another approach is discussed in \cite{BTTV}, which only quantises a maximally commuting subset of vertex simplicity constraints and thus closely resembles the gauge unfixing procedure. A benchmark which a successful proposal for the simplicity constraints in the canonical theory have to satisfy is anomaly-freeness with respect to the action of the Hamiltonian constraint. More precisely, the Hamiltonian constraint should leave the kernel of the simplicity constraints invariant. At present, none of the proposed treatments is entirely satisfactory and they deserve further research. We refer to our companion papers \cite{BTTIII, BTTV} for details.

\section{Conclusion}

In this paper we extended the new connection formulation of General Relativity in all spacetime
dimensions $D+1\ge 3$ \cite{BTTI,BTTII,BTTIII} to arbitrary gauge boson (for compact structure groups) and standard
fermionic couplings. We did not discuss Higgs fields or other scalars but remark that the 
strategy for background independent quantisation of Higgs field coupled to General Relativity 
developed for  $3+1$ dimensions \cite{ThiemannQSD5} straightforwardly generalises to higher dimensions.
Fermionic fields deserve special attention when we switch from the Lorentzian to the 
Euclidean Clifford algebra which is a necessary step if we want to maintain a compact 
structure group for General Relativity in order to have access to background independent
Hilbert space representations \cite{AshtekarQuantizationOfDiffeomorphism}. In this paper we showed that while there arise 
corresponding correction terms, there is no problem in principle in performing this switch 
and one can apply the quantisation techniques developed for $3+1$ dimensions. With this work,
the stage is set to study Supergravity theories, which we pursue in  \cite{BTTVI, BTTVII}.\\
\\
\\
{\bf\large Acknowledgements}\\
NB and AT thank Christian Fitzner for discussions about fermionic variables and the German National Merit Foundation for financial support. The part of the research performed at the Perimeter Institute for Theoretical Physics was supported in part by funds from the Government of Canada through NSERC and from the Province of Ontario through MEDT. 
During final improvements of this work, NB was supported by the NSF Grant PHY-1205388 and the Eberly research funds of The Pennsylvania State University.\\


\begin{appendix}

\section{Gamma Matrices}

\label{appendix_gamma}

The properties of the gamma matrices can be found in most textbooks on quantum field theory, see, for instance, \cite{weinberg-book1}. Their basic property is the Clifford algebra
\be  \{ \gamma^I, \gamma^J \} = 2 \eta^{IJ} \text{,}  \ee 
where $\eta^{IJ}$ is the flat Minkowski metric of a spacetime with signature $(p,q)$. 
From this relation alone, one deduces,
\be  [ \Sigma^{IJ}, \gamma^K ] = -i \gamma^I \eta^{JK} + i \gamma^J \eta^{IK}  \ee 
and 
\be  i [\Sigma^{IJ}, \Sigma^{KL}] = \eta^{LJ} \Sigma^{KI} - \eta^{LI} \Sigma^{KJ} + \eta^{JK} \Sigma^{IL} - \eta^{IK} \Sigma^{JL} \text{,} \ee 
where $\Sigma^{IJ} := -\frac{i}{4}[\gamma^I, \gamma^J]$. $\Sigma^{IJ}$ thus constitutes a representation of the Lie algebra so$(p,q)$ on spinor space. 

Furthermore, the expression $\{\gamma^K, \Sigma^{IJ}\} = -i \gamma^{[K} \gamma^I \gamma^{J]}$ is completely antisymmetric in $I, J, K$. 

It is noteworthy that $\Sigma^{IJ}$ is a Hermitian matrix for Euclidean signature. In general, $\left( \Sigma^{IJ} \right)^\dagger =\eta^{II} \eta^{JJ} \Sigma^{IJ}$, which becomes important when dealing with Lorentzian signature, i.e. the boost part of the Gau{\ss} constraint is purely rotational as $\Sigma^{0i} + (\Sigma^{0i})^\dagger = 0$.  

Explicit representations of the gamma matrices exist for all dimensions $D+1 \geq 2$, see, for instance, \cite{penrose-book2}, or \cite{ortin-book}. A generalisation of left- and right-handed spinors exists for $D+1$ even and is spelled out in our companion paper \cite{BTTVI}.

\end{appendix}

\newpage

\bibliography{pa90pub.bbl}

\providecommand{\href}[2]{#2}\begingroup\raggedright\begin{thebibliography}{10}

\bibitem{BTTI}
N.~Bodendorfer, T.~Thiemann, and A.~Thurn, ``{New variables for classical and
  quantum gravity in all dimensions: I. Hamiltonian analysis},'' {\em Classical
  and Quantum Gravity} {\bf 30} (2013) 045001, {\tt arXiv:1105.3703 [gr-qc]}.

\bibitem{BTTII}
N.~Bodendorfer, T.~Thiemann, and A.~Thurn, ``{New variables for classical and
  quantum gravity in all dimensions: II. Lagrangian analysis},'' {\em Classical
  and Quantum Gravity} {\bf 30} (2013) 045002, {\tt arXiv:1105.3704 [gr-qc]}.

\bibitem{BTTIII}
N.~Bodendorfer, T.~Thiemann, and A.~Thurn, ``{New variables for classical and
  quantum gravity in all dimensions: III. Quantum theory},'' {\em Classical and
  Quantum Gravity} {\bf 30} (2013) 045003, {\tt arXiv:1105.3705 [gr-qc]}.

\bibitem{ThiemannQSD5}
T.~Thiemann, ``{Quantum spin dynamics (QSD) V: Quantum Gravity as the Natural
  Regulator of Matter Quantum Field Theories},'' {\em Classical and Quantum
  Gravity} {\bf 15} (1998) 1281--1314, {\tt arXiv:gr-qc/9705019}.

\bibitem{KibbleCanonicalVariablesFor}
T.~W.~B. Kibble, ``{Canonical Variables for the Interacting Gravitational and
  Dirac Fields},'' {\em Journal of Mathematical Physics} {\bf 4} (1963), no.~11
  1433--1437.

\bibitem{AshtekarNewVariablesFor}
A.~Ashtekar, ``{New Variables for Classical and Quantum Gravity},'' {\em
  Physical Review Letters} {\bf 57} (1986) 2244--2247.

\bibitem{AshtekarInclusionOfMatter}
A.~Ashtekar, J.~Romano, and R.~Tate, ``{New variables for gravity: Inclusion of
  matter},'' {\em Physical Review D} {\bf 40} (1989) 2572--2587.

\bibitem{PerezPhysicalEffectsOf}
A.~Perez and C.~Rovelli, ``{Physical effects of the Immirzi parameter in loop
  quantum gravity},'' {\em Physical Review D} {\bf 73} (2006) 044013, {\tt
  arXiv:gr-qc/0505081}.

\bibitem{BojowaldCanonicalGravityWith}
M.~Bojowald and R.~Das, ``{Canonical gravity with fermions},'' {\em Physical
  Review D} {\bf 78} (2008) 064009, {\tt arXiv:0710.5722 [gr-qc]}.

\bibitem{MercuriFermionsInThe}
S.~Mercuri, ``{Fermions in the Ashtekar-Barbero connection formalism for
  arbitrary values of the Immirzi parameter},'' {\em Physical Review D} {\bf
  73} (2006) 084016, {\tt arXiv:gr-qc/0601013}.

\bibitem{DateTopologicalInterpretationOf}
G.~Date, R.~Kaul, and S.~Sengupta, ``{Topological interpretation of
  Barbero-Immirzi parameter},'' {\em Physical Review D} {\bf 79} (2009) 044008,
  {\tt arXiv:0811.4496 [gr-qc]}.

\bibitem{ThiemannKinematicalHilbertSpaces}
T.~Thiemann, ``{Kinematical Hilbert spaces for Fermionic and Higgs quantum
  field theories},'' {\em Classical and Quantum Gravity} {\bf 15} (1998)
  1487--1512, {\tt arXiv:gr-qc/9705021}.

\bibitem{AshtekarRepresentationTheoryOf}
A.~Ashtekar and J.~Lewandowski, ``{Representation Theory of Analytic Holonomy
  C* Algebras},'' in {\em Knots and Quantum Gravity} (J.~Baez, ed.), (Oxford),
  Oxford University Press1994.
\newblock {\tt arXiv:gr-qc/9311010}.

\bibitem{AshtekarDifferentialGeometryOn}
A.~Ashtekar and J.~Lewandowski, ``{Differential geometry on the space of
  connections via graphs and projective limits},'' {\em Journal of Geometry and
  Physics} {\bf 17} (1995) 191--230, {\tt arXiv:hep-th/9412073}.

\bibitem{AshtekarProjectiveTechniquesAnd}
A.~Ashtekar and J.~Lewandowski, ``{Projective techniques and functional
  integration for gauge theories},'' {\em Journal of Mathematical Physics} {\bf
  36} (1995) 2170--2191, {\tt arXiv:gr-qc/9411046}.

\bibitem{MarolfOnTheSupport}
D.~Marolf and J.~M. Mour\~{a}o, ``{On the support of the Ashtekar-Lewandowski
  measure},'' {\em Communications in Mathematical Physics} {\bf 170} (1995)
  583--605, {\tt arXiv:hep-th/9403112}.

\bibitem{AshtekarQuantizationOfDiffeomorphism}
A.~Ashtekar, J.~Lewandowski, D.~Marolf, J.~M. Mour\~{a}o, and T.~Thiemann,
  ``{Quantization of diffeomorphism invariant theories of connections with
  local degrees of freedom},'' {\em Journal of Mathematical Physics} {\bf 36}
  (1995) 6456--6493, {\tt arXiv:gr-qc/9504018}.

\bibitem{WielandComplexAshtekarVariables}
W.~Wieland, ``{Complex Ashtekar variables and reality conditions for Holst's
  action},'' {\em Annales Henri Poincar\'{e}} {\bf 13} (2012), no.~3 425--448,
  {\tt arXiv:1012.1738 [gr-qc]}.

\bibitem{BTTV}
N.~Bodendorfer, T.~Thiemann, and A.~Thurn, ``{On the implementation of the
  canonical quantum simplicity constraint},'' {\em Classical and Quantum
  Gravity} {\bf 30} (2013) 045005, {\tt arXiv:1105.3708 [gr-qc]}.

\bibitem{HenneauxQuantizationOfGauge}
M.~Henneaux and C.~Teitelboim, {\em {Quantization of Gauge Systems}}.
\newblock Princeton University Press, 1994.

\bibitem{JacobsonFermionsInCanonical}
T.~Jacobson, ``{Fermions in canonical gravity},'' {\em Classical and Quantum
  Gravity} {\bf 5} (1988) L143--L148.

\bibitem{Choquet-BruhatAnalysisManifoldsAnd2}
Y.~Choquet-Bruhat and C.~DeWitt-Morette, {\em {Analysis, Manifolds and Physics.
  Part II - Revised and Enlarged}}.
\newblock North Holland, Amsterdam, 2000.

\bibitem{BTTVI}
N.~Bodendorfer, T.~Thiemann, and A.~Thurn, ``{Towards loop quantum supergravity
  (LQSG): I. Rarita-Schwinger sector},'' {\em Classical and Quantum Gravity}
  {\bf 30} (2013) 045006, {\tt arXiv:1105.3709 [gr-qc]}.

\bibitem{FreidelBFDescriptionOf}
L.~Freidel, K.~Krasnov, and R.~Puzio, ``{BF description of higher-dimensional
  gravity theories},'' {\em Advances in Theoretical and Mathematical Physics}
  {\bf 3} (1999) 1289--1324, {\tt arXiv:hep-th/9901069}.

\bibitem{BTTVII}
N.~Bodendorfer, T.~Thiemann, and A.~Thurn, ``{Towards loop quantum supergravity
  (LQSG): II. p -form sector},'' {\em Classical and Quantum Gravity} {\bf 30}
  (2013) 045007, {\tt arXiv:1105.3710 [gr-qc]}.

\bibitem{weinberg-book1}
S.~Weinberg, {\em {The Quantum Theory of Fields, Volume 1: Foundations}}.
\newblock Cambridge University Press, 2005.

\bibitem{penrose-book2}
R.~Penrose and W.~Rindler, {\em {Spinors and Space-Time: Volume 2, Spinor and
  Twistor Methods in Space-Time Geometry}}.
\newblock Cambridge University Press, 1988.

\bibitem{ortin-book}
T.~Ort\'{\i}n, {\em {Gravity and Strings}}.
\newblock Cambridge University Press, 2007.

\end{thebibliography}\endgroup

\end{document}